\documentclass[webconf, onecolumn]{svjour}
\usepackage{ifpdf}
\ifpdf
  \usepackage[pdftex]{graphicx,color}
\else
  \usepackage[dvips]{graphicx,color}
\fi

\usepackage[varg]{txfonts} 
\usepackage[latin1]{inputenc}
\session-title{AO for ELT}
\begin{document}
\title{EAGLE Spectroscopy of Resolved Stellar Populations Beyond the Local Group}
\author{Chris Evans\inst{1}\fnmsep\thanks{\email{chris.evans@stfc.ac.uk}} \and Yanbin Yang\inst{2}
\and Mathieu Puech\inst{2} \and Matthew Lehnert\inst{2} \and Michael Barker\inst{3} 
\and Annette Ferguson\inst{3} \and Jean-Gabriel Cuby\inst{4} \and Simon Morris\inst{5}
\and G\'{e}rard Rousset\inst{6} \and Fran\c{c}ois Ass\'{e}mat\inst{6} 
\and Hector Flores\inst{2}}
\institute{UK ATC, Royal Observatory Edinburgh, Blackford Hill, Edinburgh, EH9 3HJ, UK
\and
GEPI, Observatoire de Paris, 5 Place Jules Janssen, 92195 Meudon Cedex, France
\and
Institute for Astronomy, Royal Observatory Edinburgh, Blackford Hill, Edinburgh, EH9 3HJ, UK
\and 
LAM, OAMP, 38 rue Fr\'{e}d\'{e}ric Joliot Curie, 13388 Marseille Cedex 13, France
\and
Department of Physics, Durham University, South Road, Durham, DH1 3LE, UK
\and
LESIA, Observatoire de Paris, 5 Place Jules Janssen, 92195 Meudon Cedex, France
}

\abstract{Valuable insights into galaxy evolution can be gleaned from studies of resolved
stellar populations in the local Universe. Deep photometric surveys
have provided tracers of the star-formation histories in galaxies from
0.8-16~Mpc, but without robust chemical abundances and stellar
kinematics from spectroscopy, their sub-structures and assembly
histories remain hidden from us. In this context, we introduce the
EAGLE design study for a multi--integral-field-unit, near-infrared
spectrograph for the European Extremely Large Telescope (E-ELT).
\mbox{EAGLE} will exploit the unprecedented light-gathering power of the
E-ELT to deliver AO-corrected spectroscopy across a large (38.5
arcmin$^2$) field, truly revolutionising our view of stellar
populations in the Local Volume.}

\maketitle

\section{Introduction}\label{intro}
Discoveries of disrupted satellite galaxies have
demonstrated that our evolutionary picture of the Milky Way is far
from complete \cite{m03}, let alone our understanding of galaxies elsewhere in
the Universe. Deep imaging from ground-based telescopes and the {\it
Hubble Space Telescope (HST)} has yielded colour-magnitude diagrams
(CMDs) with unprecedented fidelity, providing new and exciting views
of the outer regions of galaxies beyond the Milky Way for the first
time, e.g. in M31 \cite{f05,rich08} and M33 \cite{b07}.  From comparison
with stellar evolutionary models, these data are used to explore the
star-formation and chemical-enrichment histories of the targeted
regions, providing a probe of the past evolution and, in particular,
the merger/interaction histories of these external galaxies. There is
increasing evidence for the accretion of numerous low-mass satellite
galaxies in the assembly of the present-day Milky Way \cite{b06}.  Do
we see evidence for similar processes at work in other large galaxies? 
Moreover, what are the assembly histories in galaxies with very
different morphological types, such as massive ellipticals, large
metal-poor irregulars, and lower-mass, late-type spirals like M33?

Photometric methods are immensely powerful when applied to
extragalactic stellar populations, but only via precise chemical
abundances and stellar kinematics can we break the age-metallicity
degeneracy, while also disentangling the populations associated with
different structures, i.e. follow-up spectroscopy is required.  Over
the past decade the Calcium Triplet (CaT, spanning 0.85-0.87 $\mu$m)
has become an increasingly used diagnostic of stellar metallicities
and radial velocities in nearby galaxies, providing new views of their
star-formation histories and sub-structure, e.g. the Dwarf Abundances
and Radial velocities Team (DART) large programme with VLT-FLAMES
\cite{t04}.  However, 8-10m class telescopes are already at their
limits in pursuit of CaT spectra of the evolved populations in
galaxies at distances greater than $\sim$300~kpc,
e.g. Keck-DEIMOS observations in M31 struggled to yield useful
signal-to-noise below the tip of the red giant branch at $I$~$>$~21.5 \cite{c06}.

\section{Resolved Stellar Populations in the ELT Era}

With its vast primary aperture and excellent angular resolution, the
E-ELT will be {\em the} facility to unlock spectroscopy of evolved
stellar populations in the broad range of galaxies in the Local
Volume, from the edge of the Local Group, out towards the Virgo
Cluster. This will bring a wealth of new and exciting target galaxies
within our grasp, spanning a broader range of galaxy morphologies,
star-formation histories and metallicities than those available to us
at present in the Local Group.  These observations can then be used to
confront theoretical models to provide a unique view of galaxy
assembly and evolution.  There are many compelling and ground-breaking
targets for stellar spectroscopy with the E-ELT including, in order of distance:
\begin{itemize}
\item{NGC 3109 and Sextans A with sub-SMC metallicities ($Z < 0.2 Z_\odot$), both at 1.3 Mpc.}
\item{The spiral dominated Sculptor `Group' at 2-4 Mpc.}
\item{The M83/NGC5128 (Centaurus A) grouping at $\sim$4-5 Mpc.}
\item{NGC 3379, the nearest normal elliptical at 10.8 Mpc.}
\item{The Virgo Cluster of galaxies at 16-17 Mpc, the nearest massive cluster.}
\end{itemize}

In contrast to proposed E-ELT observations of high-redshift galaxies,
targets for CaT spectroscopy are readily available.  For example, deep
ground-based and {\it HST} imaging in galaxies in the Local Volume has
begun to investigate their stellar populations
\cite{r05,ghosts,angst}, yet the stellar magnitudes are well beyond
spectroscopy with existing facilities.  Note that although we have
focussed mostly on southern-hemisphere targets here, there are equally compelling
northern hemisphere targets, including the M81 group, and deeper
studies in M31 and M33.

\section{EAGLE: A Multi-IFU, Near-IR Spectrograph for the E-ELT}

\begin{figure}
\begin{center}
\includegraphics{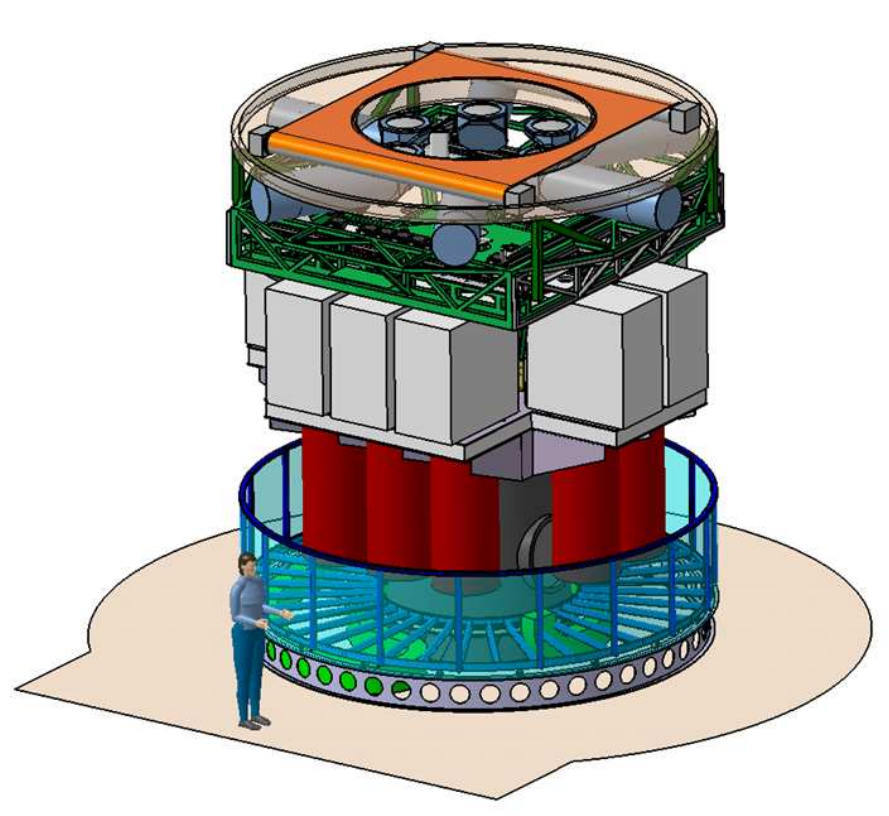}
\caption{Baseline mechanical design of EAGLE.}\label{eagle}
\end{center}
\end{figure}

The EAGLE Phase A study is a French-UK partnership to provide an
advanced conceptual design of an AO-corrected, near-infrared
spectrograph with multiple integral-field units (IFUs). The
specifications of the baseline design are summarised in
Table~\ref{specs}, with the mechanical implementation shown in
Figure~\ref{eagle}.  To minimise flexure on the instrument over the
course of each night, EAGLE is intended to be located at the
gravity-invariant focus of the E-ELT.

\begin{table}[h]
\begin{center}
\caption{EAGLE Baseline Design. The patrol field is the instrument field-of-view
within which IFUs can be configured to observe individual targets.}\label{specs}       
\begin{tabular}{ll}
\hline\noalign{\smallskip}
Parameter & Specification \\
\noalign{\smallskip}\hline\noalign{\smallskip}
Patrol Field & Eqv. 7$^\prime$ diameter \\
IFU field-of-view & 1{\mbox{\ensuremath{.\!\!^{\prime\prime}}}}65 $\times$ 1{\mbox{\ensuremath{.\!\!^{\prime\prime}}}}65 \\
Multiplex (\# of IFUs) & 20 \\
Spatial resolution & 30\%EE in 75 mas ($H$ band)\\
Spectral resolving power ($R$) & 4,000 \& 10,000\\
Wavelength range & 0.8-2.5$\mu$m\\
\noalign{\smallskip}\hline
\end{tabular}
\end{center}
\end{table}

\noindent Five top-level science topics have shaped the
EAGLE design:
\begin{itemize}
\item{Physics and evolution of high-redshift galaxies}
\item{Detection and characterisation of `first light' galaxies}
\item{Galaxy assembly and evolution from stellar archaeology}
\item{Star-formation, stellar clusters and the initial mass function}
\item{Co-ordinated growth of black holes and galaxies}
\end{itemize}
A summary of the high-redshift case and of E-ELT observations in the
Galactic Centre region (combining elements of the last two) are given
elsewhere in these proceedings \cite{p09,tp09}.  Here we focus on 
simulated EAGLE performances for spectroscopy of resolved stellar
populations, used to probe the assembly and evolution of their host galaxies.

\section{EAGLE Adaptive Optics}
EAGLE will employ multi-object adaptive optics (MOAO) to provide
significantly improved image quality for selected target fields within
the focal plane \cite[for example]{falcon}.  The EAGLE baseline design
uses an array of six laser guide stars and five natural guide stars
(NGS) to map the atmospheric turbulence.  The deformable mirror in the
telescope (M4) will be used to correct for the low-order wavefront
error terms, with the high-order terms corrected by deformable mirrors
in each science channel. An integral part of the EAGLE project is the
CANARY on-sky demonstrator on the William Herschel Telescope in La
Palma, discussed elsewhere in these proceedings \cite{m09}.

The consortium has undertaken extensive simulations of MOAO
performance, paying attention to the location and magnitude of
potential NGS in real example fields \cite{r09}.  Here we investigate
the scientific peformance of EAGLE, making use of simulated
point-spread functions (PSFs) from two NGS configurations (`\#10' and
`\#1'), which are illustrative of relatively good and poor performance given
the spatial distribution and magnitude of the available guide stars
\cite{r09}.

\section{EAGLE Performance: CaT Spectroscopy}

We have investigated EAGLE's performance in the CaT region,
incoporating the new PSFs, using a modified version of the IFU
simulation tool developed to characterise the MOAO requirements for
observations of high-redshift galaxies
\cite{p08}. The principal assumptions in the CaT simulations were:
\begin{itemize}
\item{Telescope: 42-m primary, with central obscuration of 9\%}
\item{Exposure time: 10 hrs (20x1800s)}
\item{$\lambda$-range: 8400-8750 \AA, and $R$ = 10,000}
\item{Spatial sampling: 37.5 mas spatial pixels}
\item{Total throughput: telescope (0.8), atmosphere (0.95), instrument \& detector (0.25 @ 0.85$\mu$m): 0.19}
\item{Detector read-out noise: 5e$^-$/pixel}
\item{Dark current: 0.01e$^-$/pixel/s}
\end{itemize}

An example CaT spectrum (T$_{\rm eff}$ = 4,000\,K, log{\it g} = 2.0)
is adopted as a template from the synthetic spectra calculated using
Kurucz model atmospheres for the {\it GAIA} mission \cite{mc00}.  The
synthetic spectra were calculated for $R=$20,000, i.e. sufficiently
over-sampled so as to be degraded to either of the two spectral
resolving powers provided by EAGLE.  To this template we append an
additional continuum region to provide a line-free region with which
to investigate to the signal-to-noise (S/N) of the spectra in the
resulting datacubes.

Assuming a Paranal-like site, we have investigated two sets of seeing
conditions in the simulations:
0{\mbox{\ensuremath{.\!\!^{\prime\prime}}}}65 at
$\lambda$\,=\,0.5$\mu$m at zenith (the mean VLT seeing at Paranal, \cite{sm08})
and 0{\mbox{\ensuremath{.\!\!^{\prime\prime}}}}90 at a zenith distance
(ZD) of 35$^\circ$, providing a good investigation of the performance
from execution of a `Large Programme'-like survey.
The S/N results for the two NGS configurations are given in
Table~\ref{sims1}.  These results are for spectra extracted from the
central spatial pixel of a point source at the centre of the cube
(optimal PSF-fitting extractions should be able to improve on these).
The key result is that {\it S/N\,$\ge$\,10 is recovered from a stacked
10\,hr exposure at I\,=\,24.5, in mean seeing, in both NGS
configurations}.  This corresponds to spectroscopy of stars at the tip of the 
red giant branch (RGB), with M$_I$\,=\,$-$4, out to $\sim$5\,Mpc in just 10\,hrs.
This is four magnitudes deeper than FLAMES-GIRAFFE using the LR08
grating ($R$\,$=$\,6,500) with the same exposure time.  

\begin{table}[h]
\begin{center}
\caption{EAGLE CaT results: Continuum S/N obtained for $R$ = 10,000, $t_{\rm exp}$ = 10 hrs. }\label{sims1}       
\begin{tabular}{ccccc}
\hline\noalign{\smallskip}
& \multicolumn{2}{c}{Seeing = 0{\mbox{\ensuremath{.\!\!^{\prime\prime}}}}9 @ ZD=35$^\circ$} &
\multicolumn{2}{c}{Mean VLT Seeing (0{\mbox{\ensuremath{.\!\!^{\prime\prime}}}}65 @ ZD=0$^\circ$)} \\
$I_{\rm VEGA}$ & NGS `good' & NGS `poor' & NGS `good' & NGS `poor' \\
\noalign{\smallskip}\hline\noalign{\smallskip}
22.5 &  40 &  27 & 56 & 48 \\
23.5 &  16 &  11 &  28 & 24 \\
24.5 &  $\phantom{1}$8 & $\phantom{1}$4 & 13 & 10 \\
\noalign{\smallskip}\hline
\end{tabular}
\end{center}
\end{table}

Some high-order contributions to the wavefront error, as well as
contributions from tip-tilt and defocus, are not included in the
simulated PSFs \cite{r09}.  Scaling the PSFs for this in science
simulations is non-trivial as the flux will be conserved
(re-distributed) rather than lost.  
In additional simulations the PSFs were artificially scaled by a further
throughput factor (0.65) to give a pessimistic example for a given
turbulent profile, seeing etc., still yielding S/N\,$\sim$\,10 for $I$\,=\,24.5
in the good NGS configuation, in mean seeing, in 10\,hrs

Although we have focussed on the $R$~=~10,000 case here, similar
calculations at $R$~=~4,000 for $I$\,=\,24.5 and 26.0 (with the latter corresponding
to approximately the tip of the RGB in NGC\,3379) yield
S/N\,$\ge$\,10 in 5 and 80\,hrs, respectively.

\section{EAGLE Observations in the Sculptor Group}

An example Large Programme for EAGLE is to undertake spectroscopy of
evolved stars in the galaxies in the Sculptor Group, comprising five
spirals (NGC\,55, NGC\,247, NGC\,253, NGC\,300 and NGC\,7793) and
numerous dwarf irregulars.  Distance estimates over the past decade
have revealed that this `group' is actually two distinct components \cite{k03}, 
at approximately 1.9\,Mpc (NGC\,55 \& 300) and 3.6-3.9\,Mpc (NGC\,247, 253 \& 7793).
These five galaxies represent {\em the most immediate opportunity to
study the star-formation history and mass assembly of spirals} beyond
the limited sample available at present, i.e. the Milky Way, M31 and
M33.  Their masses are in the range 1.5-8$\times$10$^{10}$\,M$_\odot$,
putting them on a par with M33 -- it is exactly these late-type,
low-mass, small bulge (or even bulge-less) spirals which are the
systems that theoretical N-body/semi-analytic simulations struggle
hardest to reproduce \cite{db04}.

\begin{figure}
\begin{center}
\begin{tabular}{cc}
\resizebox{0.48\columnwidth}{!}{\includegraphics{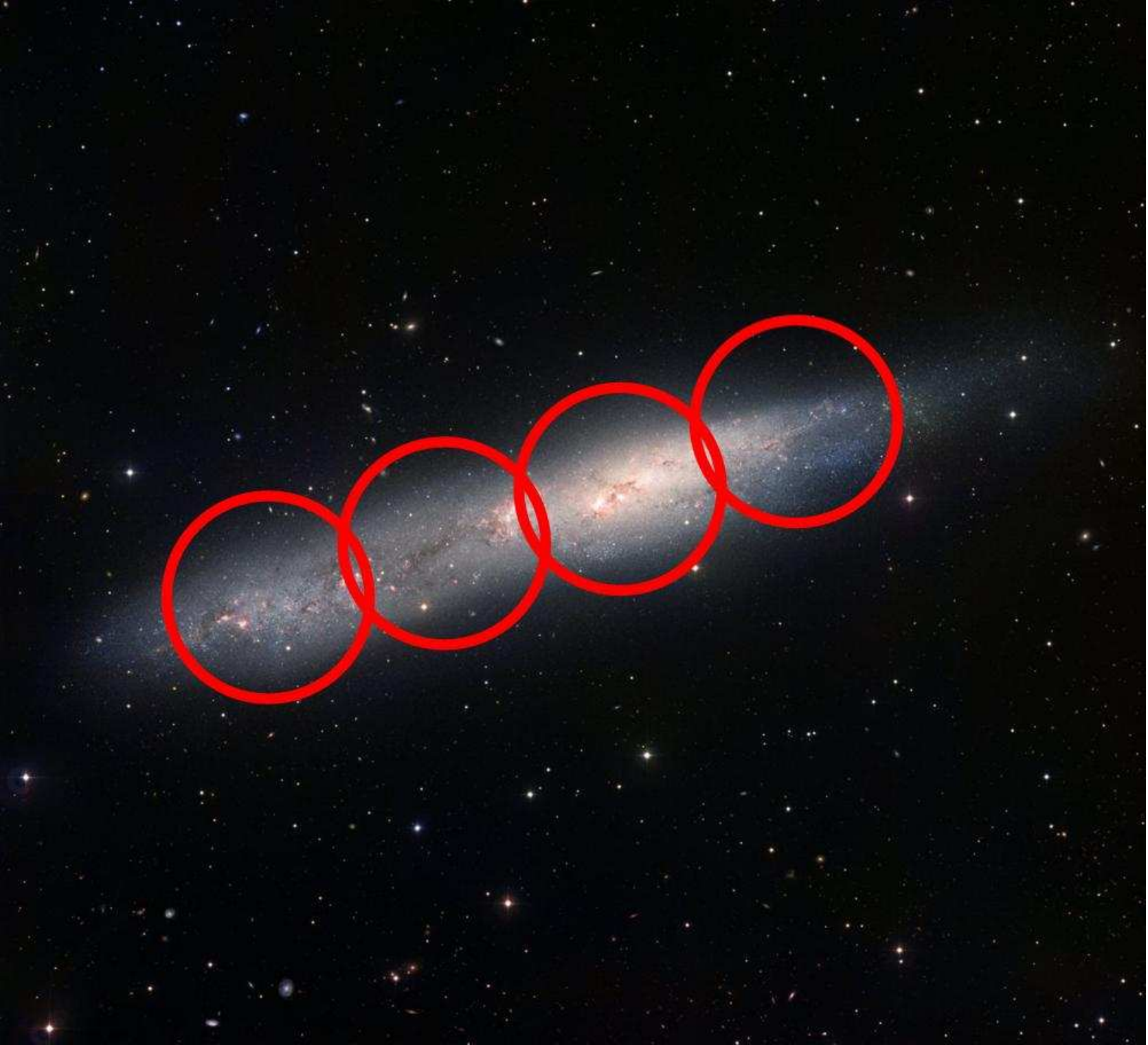}} & \resizebox{0.483\columnwidth}{!}{\includegraphics{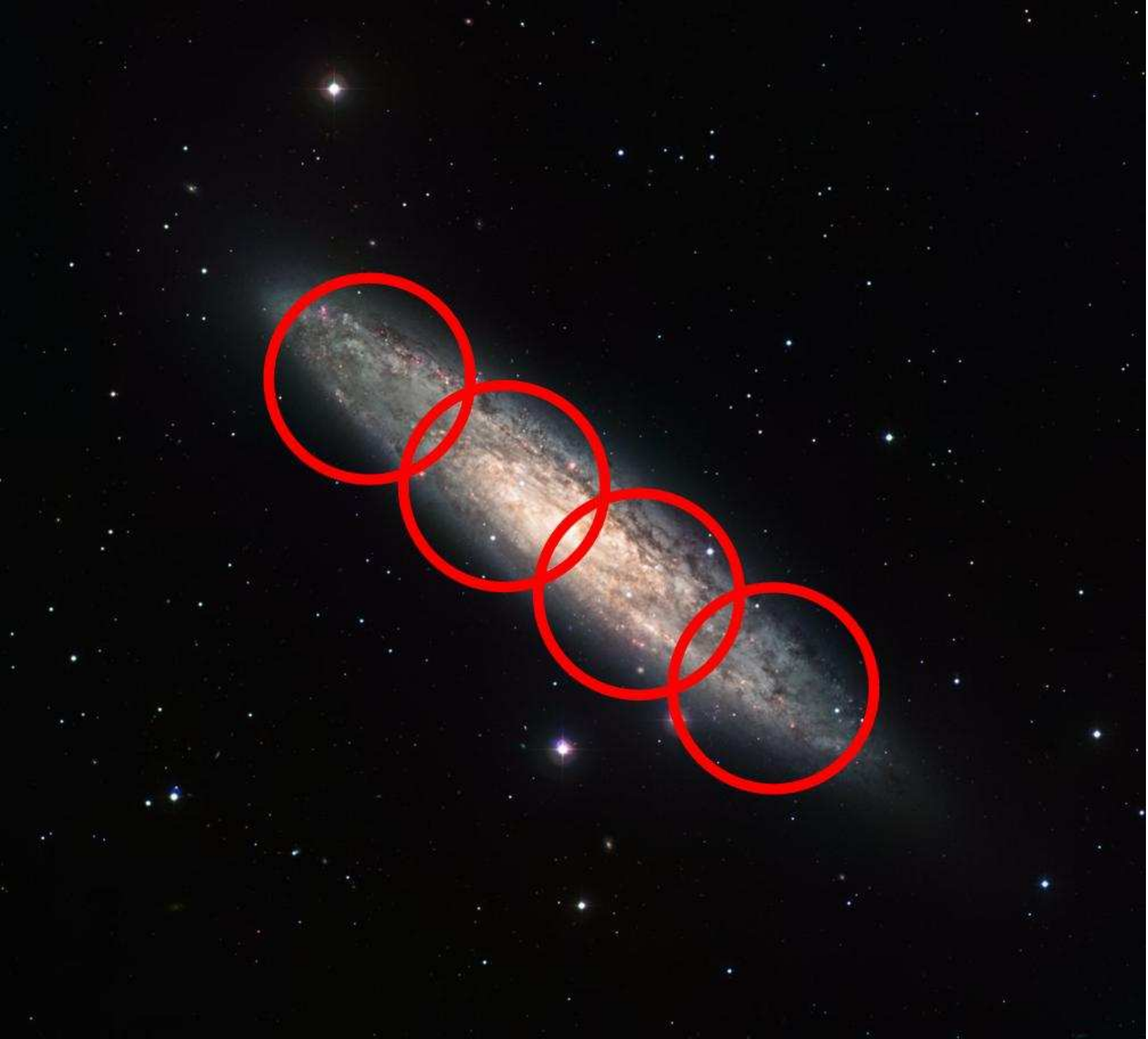}}
\end{tabular}
\caption{Example major-axis EAGLE pointings in NGC\,55 at 1.9\,Mpc ({\it left}) and NGC\,253 at 3.6\,Mpc
({\it right}). The central (unvignetted) 5 arcmin field of EAGLE is illustrated by the
red circles. [Original image credits: ESO].}\label{55_253}
\end{center}
\end{figure}

We propose spectroscopy of stars in the upper RGB, spanning
M$_I$\,=\,$-$4 (at the tip of the RGB) to approximately
M$_I$\,=\,$-$2, i.e. 22.5\,$<$\,$I$\,$<$24.5 in NGC\,55 and NGC\,300,
and down to $I$\,$\sim$\,24.5 (i.e. M$_I$\,$\sim$\,$-$3) in the
others.  Targets would be observed along the major and minor axes of
each galaxy, sampling the stellar populations across different spiral
structures and the halos.  For instance, NGC\,55 and NGC\,253 both
have semi-major axes greater than ten arcminutes, i.e. multiple EAGLE
pointings are required to survey stars across the full extent of such
a galaxy, as shown in Figure~\ref{55_253}.  In all but the most
rarefied regions, multiple stars will be observed with each IFU (thus
boosting the effective multiplex), with the IFUs also enabling better
sky subtraction compared to fibre/slit methods.  Adopting
10\,hrs/pointing for NGC\,55/NGC\,300 and 15\,hrs/pointing for the
other three galaxies (cf. the sensitivity results in
Table~\ref{sims1}), well in excess of 1,000 stars can be observed in a
total of $\sim$240 hrs.

\begin{figure}
\begin{center}
\begin{tabular}{cc}
\resizebox{0.38\columnwidth}{!}{\includegraphics{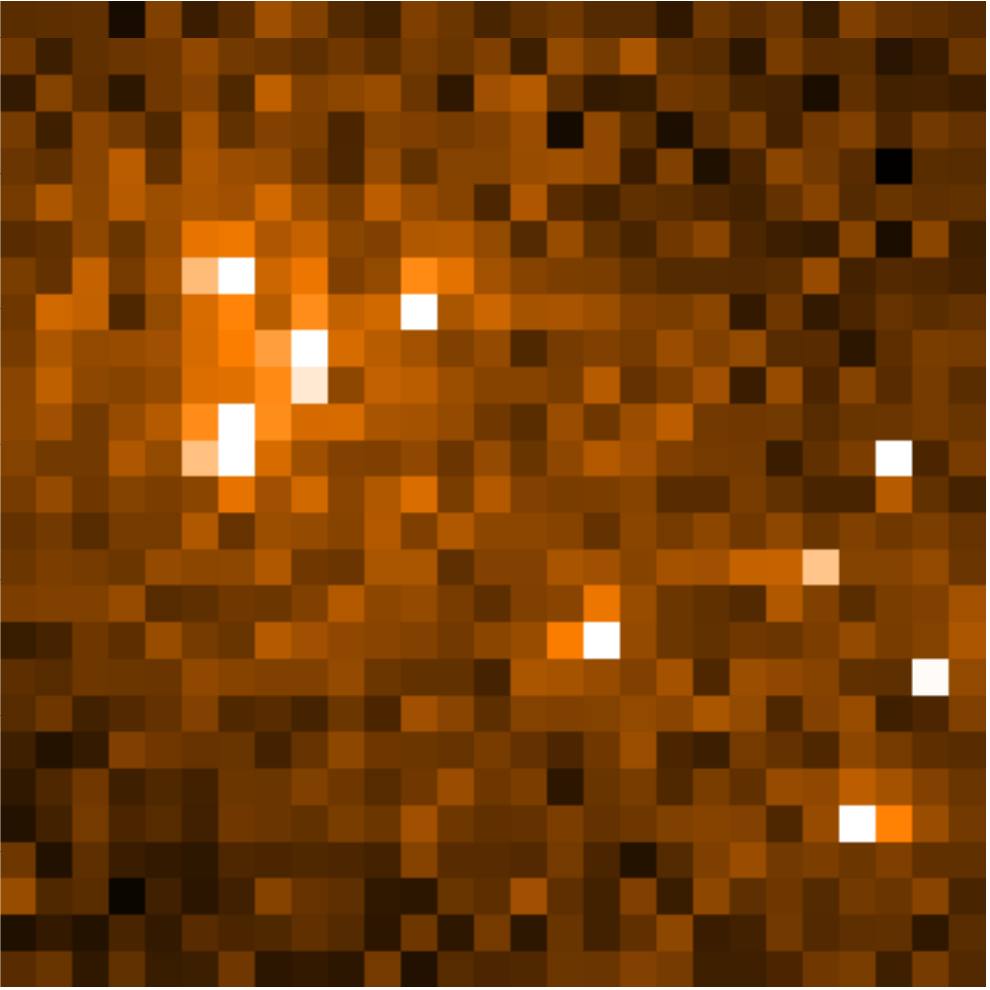}} & \resizebox{0.58\columnwidth}{!}{\includegraphics{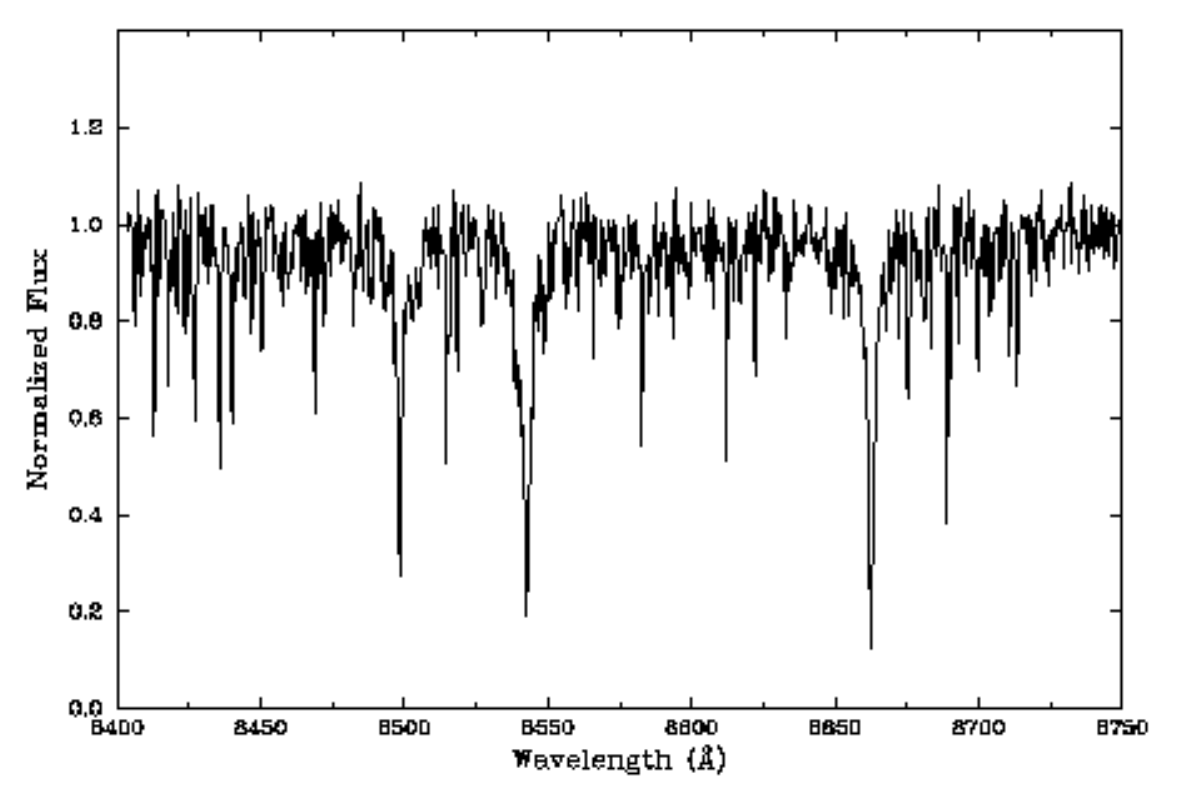}}
\end{tabular}
\caption{{\it Left:} Front slice of simulated IFU datacube for EAGLE observations in the central region of NGC\,55; 
{\it Right:} Simulated CaT spectrum for a star with $I$ = 23.5, yielding S/N $\sim$ 25 in the continuum.}\label{data}
\end{center}
\end{figure}

The left-hand panel of Figure~\ref{data} shows a typical
1$^{\prime\prime}\,\times\,$1$^{\prime\prime}$ IFU pointing in the
central region of NGC\,55 (in which the simulations adopted a
smaller IFU than in the baseline design simply to reduce the
computation time).  The magnitudes and relative positions of the stars
are from {\it HST} Advanced Camera for Surveys (ACS) observations in
the core region of NGC\,55, taken as part of the GHOSTS survey
\cite{ghosts}.  This example illustrates perfectly the gain in
effective multiplex from the IFUs (i.e. nine stars in this
1$^{\prime\prime}\,\times\,$1$^{\prime\prime}$ region), with minimal 
impact from crowding.  A simulated
CaT spectrum ($I$\,=\,23.5, S/N\,$\sim$\,25) is shown in the
right-hand panel of Figure~\ref{data}.

Recent simulations predict that stellar radial mixing (also called
radial `churning', or orbit switching) due to perturbations from
transient spiral density waves plays a large role in shaping the age
and metallicity gradients \cite{r08} -- this dynamical process could have huge
implications for stellar archaeology, as it may modify, or even erase,
the original gradients.  Ages and metallicities will be derived for
each star to investigate the gradients of these properties across each
galaxy, while also inspecting the results for evidence of
sub-structure.  In outer halo fields we will also test the prediction
that halo stellar metallicity is thought to scale with halo stellar
mass \cite{m05,f06}.

\vspace*{-0.1in}
\section{Summary}
The EAGLE design provides a unique combination of abilities to 
harness the power of the E-ELT for spectroscopy of resolved stellar
populations.  The image quality from MOAO will be significantly
better than that obtained from seeing-limited or ground-layer AO modes,
enabling us to explore spatially-resolved, extragalactic stars across
wide fields of over five arcminutes.

EAGLE spectroscopy in the Sculptor Group is just one illustration of a
potential Large Programme.  Even within the Sculptor galaxies one
might prioritize, e.g., NGC\,253 (due to its starburst activity) or
NGC\,300 (due to its relative proximity) first, building-up a
comprehensive survey in one galaxy or following-up on interesting
sub-structures before turning to the others.  Moreover, 
deeper observations of the most luminous evolved stars 
in selected fields in galaxies at 10\,Mpc and beyond will also be hugely attractive.

\end{document}